\documentclass[aps,pra,twocolumn]{revtex4}

\usepackage{amsmath}
\usepackage{amssymb}
\usepackage{epsfig}
\usepackage{color}
\usepackage{soul}
\usepackage{epstopdf}

\graphicspath{{.}{./figs/}}

\newcommand{\braket}[2]{\langle{#1}|{#2}\rangle}

\newcommand* {\ket}[1]{\ensuremath{| {#1} \rangle}}

\begin{document}
\title{Quantum state matching of qubits via measurement-induced nonlinear transformations}

\author{Orsolya K\'{a}lm\'{a}n}
\author{Tam\'{a}s Kiss}
\affiliation{
Institute for Solid State Physics and Optics, Wigner Research Centre, Hungarian Academy of Sciences, 
P.O. Box 49, H-1525
Budapest, Hungary}
%\date{\today}
\begin{abstract}
We consider the task of deciding whether an unknown qubit state falls in a prescribed neighborhood of a reference state. We assume that several copies of the unknown state are given and apply a unitary operation pairwise on them combined with a post-selection scheme conditioned on the measurement result obtained on one of the qubits of the pair. The resulting transformation is a deterministic, nonlinear, chaotic map in the Hilbert space. We derive a class of these transformations capable of orthogonalizing nonorthogonal qubit states after a few iterations. These nonlinear maps orthogonalize states which correspond to the two different convergence regions of the nonlinear map.  
Based on the analysis of the border (the so-called Julia set) between the two regions of convergence, we show that it is always possible to find a map capable of deciding whether an unknown state is within a neighborhood of fixed radius around a desired quantum state. We analyze which one- and two-qubit operations would physically realize the scheme. It is possible to find a single two-qubit unitary gate for each map or, alternatively, a universal special two-qubit gate together with single-qubit gates in order to carry out the task. We note that it is enough to have a single physical realization of the required gates due to the iterative nature of the scheme.
\end{abstract}
\maketitle
\section{Introduction}

Advances of quantum technology may eventually enable the application of quantum machines for information storage and procession. Since the birth of quantum information science one is tempted to think of the output of such machines as qubits in some pure quantum state. However, the identification of such a quantum state may be nontrivial and unavoidably probabilistic. Conventional quantum state discrimination schemes \cite{Paris,Hayashi1,Hayashi2} offer to discriminate between two unknown states with an appropriately designed set of operations. The principle of universal programmable quantum-state discriminators offers a way to treat the problem more generally, comparing the data qubit to either two unknown states or one known and one unknown state \cite{Bergou1,Bergou2,Probst-Schendzielorz} or utilizing some other piece of information \cite{Colin2011}. Iteratively applying measurement-induced nonlinear evolution may provide a feasible, alternative approach to similar problems. It has been demonstrated on the example of a cavity assisted atomic scheme that its iterative application is capable of orthogonalizing quantum states of two-level atoms discriminating them according to the sign of the real part of their excitation amplitude, even for weakly excited atoms \cite{Torres2017}. 

We pose the following task:  Let us imagine a machine which is expected to produce some desired pure quantum state, but there is some systematic error in its operation and the resulting pure state is only close to the desired state. One would like to accept this resulting state if the error is small i.e., the distance from the prescribed state is within a given small interval, or with other words it falls in a prescribed neighborhood of the desired state. This problem is different compared to the usual questions asked in quantum state discrimination \cite{Paris} and comparison \cite{Barnett2003,Chefles2004,Jex2004,Kleinmann2005}, since here we would like to select a specific area around a reference state in the abstract Hilbert space. Perhaps the most closely related idea is quantum template matching by Sasaki, Carlini and Jozsa \cite{Sasaki2001,Sasaki2002}, where one would like to know which one from a given set of states resembles most to the unknown state. In our case, we have a single reference state, but all other states within a certain neighborhood are accepted as matched states, therefore we use the term quantum state matching.

Nonlinear quantum mechanics would allow for solving hard problems efficiently, for example, perfect state discrimination \cite{Childs2016,Abrams1998}. In standard quantum mechanics unitary evolution is linear but selective measurements may lead to nonlinear effects \cite{Filippov2017,Terno1999,Totharxiv}. Nonlinear state transformations in standard quantum mechanics can arise when identically prepared quantum systems are subjected to an entangling unitary transformation and subsequent selective measurements are performed on parts of the system \cite{Gisin}. Iterating such post-selective nonlinear quantum state transformations may result in a strong dependence on the initial conditions and in complex chaos in the dynamics \cite{Kiss2006,PaulaNeto2016,Kiss2008}.

Typical examples involving repeated sequences of unitary entangling transformations and selective conditional quantum measurements are quantum state purification protocols acting on identically prepared weakly entangled two-qubit systems  \cite{Alber2001}, as introduced by Bennett and Deutsch \cite{Bennett97,Deutsch98}. In these protocols the parameters are designed in such a way that quantum state purification can be achieved and a complicated dependence on initial conditions is avoided \cite{Macchiavello98}. Nevertheless, it has been demonstrated that in general an intricate dependence of the resulting  entanglement on the initially prepared states will occur \cite{Kiss2011,Guan2013}.  

Sensitivity to initial conditions can also be used for amplifying small initial differences of quantum states (a 'Schr\"odinger's microscope' as suggested by Lloyd and Slotine \cite{Lloyd2000}). The state difference amplification necessarily requires a large number of identically prepared systems which have to be discarded during the process, nevertheless resulting in an optimally scaling state discrimination procedure under certain conditions \cite{Gilyen2016}.

 In this paper we determine the class of nonlinear quantum state transformations capable of deciding whether a given qubit state matches a reference state with a prescribed precision. These nonlinear transformations have two attractive fixed points which correspond to two orthogonal pure quantum states. After the iteration of the nonlinear process the state of the qubit may end up in either of these two orthogonal pure states  depending on the initial state. We explore this new approach and show that using such nonlinear quantum state transformations one is able to distinguish between  sets of quantum states, i.e., discriminate quantum states that are in a predefined vicinity of a given reference state from quantum states which are further.  In this approach the matched qubits are not directly measured, but rather they are prepared in the ideal reference state in a nondemolition sense. The resulting qubit, prepared in the corrected state, can be used  for further processing, thus this procedure may also be viewed as quantum state error correction.

The paper is organized as follows. In Sec.~\ref{class} we determine how to construct -- starting from the simplest orthogonalizing superattractive nonlinear transformation -- all other such nonlinear transformations which are capable of orthogonalizing quantum states after only a few iterations. In Sec.~\ref{qstateid} we show how such nonlinear maps can be used to decide whether an unknown pure quantum state is in a prescribed neighborhood of a desired reference pure state. In Sec.~\ref{realization} we present a direct approach to implement a nonlinear map for quantum state matching by using a two-qubit unitary transformation (Sec.~\ref{directreal}), and an approach based on decomposing the map into a special, "contracting" nonlinear map (realizable with a two-qubit operation) plus a single-qubit rotation (Sec.~\ref{specialreal}). In Sec.~\ref{succprob} we present the success probability of the protocol, while in Sec.~\ref{initnoise} we consider mixed-state inputs. We conclude in Sec.~\ref{concl}. 

\section{Orthogonalizing nonlinear maps}
\label{class}

The nonlinear transformations we consider arise as a consequence of postselection based on measurement results in a two-qubit system, where initially the qubits are independent and they are in the same pure one-qubit state $\ket{\psi_{0}}$
\begin{equation}
\ket{\psi_{0}}=\frac{\ket{0}+z\ket{1}}{\sqrt{1+\left|z\right|^{2}}}, \quad (z\in\mathbb{C}).
\label{psi0}
\end{equation}
The state of the composite system is then of the form
\begin{equation}
\ket{\Psi_{0}}=\ket{\psi_{0}}_{A}\otimes\ket{\psi_{0}}_{B}.
\label{Psi12}
\end{equation}
One applies some entangling two-qubit operation on them and then measures the state of one of the qubits, say that of qubit $B$. The other qubit ($A$) is kept or discarded depending on the result of this measurement. The initial single-qubit state $\ket{\psi_{0}}_{A}$  after the postselection generally reads
\begin{equation}
\ket{\psi_{1}}_{A}\sim \ket{0}_{A}+f(z)\ket{1}_{A}\, ,
\end{equation}
where $f(z)$ is a complex quadratic rational function of $z$
\begin{equation}
f(z)=\frac{a_{0}z^2+a_{1}z+a_{2}}{b_{0}z^2+b_{1}z+b_{2}}\, .
\end{equation}
If $a_{0}$ and $b_{0}$ are not both zero and the polynomial in the nominator does not have a common root with the polynomial in the denominator, then we arrive at a genuine nonlinear transformation of the initial qubit state. In fact, to any given quadratic rational function one can construct a two-qubit unitary gate which physically realizes it with the above scheme \cite{Gilyen2016}.

If one has an ensemble of qubits in the same initial state $\ket{\psi_{0}}$, then by taking pairs of these qubits and applying the protocol on them, then forming new pairs of qubits from the postselected ones, the nonlinear map $f(z)$  may be iterated. After the $n$th step the initial qubit state will be transformed into
\begin{equation}
\ket{\psi_{n}}=\frac{\ket{0}+f^{(n)}(z)\ket{1}}{\sqrt{1+\left|f^{(n)}(z)\right|^{2}}}.
\end{equation} 
Due to the nonlinearity of the transformation $f(z)$, the resulting single-qubit state can be very sensitive to the initial conditions. Performing the protocol separately on two ensembles $I$ and $II$ in states $\ket{\psi_{I}}$ and $\ket{\psi_{II}}$, respectively, where $\ket{\psi_{I}}$ and $\ket{\psi_{II}}$ initially have a large overlap, the remaining qubits from the ensembles may end up in states with a small, or even zero overlap, depending on the properties of the nonlinear transformation $f(z)$ \cite{Torres2017}. 

The basic properties of $f(z)$ are determined by its multipliers $\mu_{i}=f\rq(z_{i})$ $(i=1,2,3)$, where $z_{i}$ are the fixed points of the map, such that $f(z_{i})=z_{i}$. If two out of the three fixed points of $f(z)$ are attractive, i.e., the corresponding multipliers are $\mu_{i}< 1$ (the third fixed point of $f(z)$ is in this case necessarily repelling \cite{MilnorGD,Milnorbook}) and these fixed points correspond to orthogonal qubit states, then one may expect that initially close pure states from the two different convergence regions become practically orthogonal after some number of iterations of the map. We will show that such maps may be of particular interest as they can be used for quantum state matching. 

Since we aim at using such nonlinear transformations as quantum informational tools, it is desirable to minimize the number of steps of the protocol. This requirement is due to the fact that the amount of resources needed for the implementation of a nonlinear map grows exponentially with the number of iterations \cite{Gilyen2016}. Therefore, in what follows we will explore the class of nonlinear maps $f(z)$ which satisfy the following requirements: their two stable fixed points are superattractive, i.e., their multipliers are $\mu_{1}=\mu_{2}=0$, and the two fixed points $z_{1}$ and $z_{2}$ correspond to orthogonal states. We will call these maps \textit{orthogonalizing superattractive (nonlinear) maps}. As it can be easily seen, the requirement of orthogonality is fulfilled if $z_{2}=-1/z_{1}^{\ast}$. 

\medskip

In order to find maps which fit the above requirements, we use the fact that the multipliers $\mu_{1}, \mu_{2}$ and $\mu_{3}$ of a quadratic rational function $f$ determine a conjugacy class of the map $f$, i.e., when conjugating $f$ with a M\"{o}bius transformation $g(z)=\dfrac{az+b}{cz+d}$ ($a,b,c,d\in\mathbb{C}$, $ad-bc\neq 0$), the multipliers of the transformed map $f\rq=g\circ f \circ g^{-1}$ are left unchanged \cite{MilnorGD,Milnorbook}. Therefore, any member of the conjugacy class of $f$ can be found by starting from the so-called fixed-point normal form \cite{MilnorGD}
\begin{equation}
f_{N}(z)=\frac{z\left(z+\mu_{1}\right)}{\mu_{2}z+1}, \quad \mu_{1}\mu_{2}\neq 1.
\label{normalform}
\end{equation}
The fixed points of $f_{N}(z)$ are $z_{1}=0$ with multiplier $\mu_{1}$ and $z_{2}=\infty$ with multiplier $\mu_{2}$. We note that the third fixed point $z_{3}$ and its multiplier $\mu_{3}$ are determined by $\mu_{1}$ and $\mu_{2}$. It can be easily shown that $z_{3}=\frac{1-\mu_{1}}{1-\mu_{2}}$, and $\mu_{3}=\frac{2-\mu_{1}-\mu_{2}}{1-\mu_{1}\mu_{2}}$ \cite{MilnorGD}. 

We are interested in finding superattractive nonlinear maps (maps with two superattractive fixed points), i.e., the conjugacy class described by the multipliers $\mu_{1}=\mu_{2}=0$, in which case Eq.~(\ref{normalform}) leads to the basic superattractive map $f_{0}(z)=z^{2}$ \cite{Gisin}. This map is an orthogonalizing transformation as well, since its fixed points $z_{1}=0$ and $z_{2}=\infty$ correspond to the orthogonal pure states $\ket{\psi_{z_{1}}}=\ket{0}$ and $\ket{\psi_{z_{2}}}=\ket{1}$, respectively. This means that after a few iterations of the map initial pure quantum states with $\left|z\right|<1$ will end up in $\ket{0}$, while initial states with $\left|z\right|>1$ will end up in $\ket{1}$. The repelling fixed point in this case is $z_{3}=1$ with multiplier $\mu_{3}=2$. This point, together with all the points for which $\left|z\right|=1$ are elements of the compact subset called the Julia set. The Julia set is a closed set and contains all the repelling fixed cycles. Outside of this set, iterates of $f_{0}$ lead to either of the attractive fixed points \cite{Milnorbook}. Since $f_{0}$ has two superattractive fixed points, its Julia set is a connected one \cite{MilnorGD}, therefore, it may be parameterized as a circle of radius 1 with its center in the origin of the complex plane ${\cal{J}}_{f_{0}}=\left\lbrace e^{i\varphi}, \varphi\in \left[0,2\pi\right)\right\rbrace$.

Our aim of finding the class of orthogonalizing superattractive maps then translates to finding those M\"{o}bius transformations which transform $f_{0}$ into any orthogonalizing superattractive map. Since the multipliers are not changed when conjugating with a M\"{o}bius transformation, our only requirement is that the fixed points be transformed keeping the property $z_{2}=-1/z_{1}^{\ast}$. As shown in Appendix~\ref{appA}, when conjugating a quadratic rational map by a  M\"{o}bius transformation the fixed points $z_{i}$ ($i=1,2,3$) are transformed according to Eq.~(\ref{fptransf}). In the case of the basic map $f_{0}$ this means that 
\begin{align}
&z_{0,1}=0 \overset{g}\longrightarrow z_{1}=\frac{b}{d} \notag \\
&z_{0,2}=\infty \overset{g}\longrightarrow z_{2}=\frac{a}{c}=-\frac{1}{z_{1}^{\ast}} \label{z1z2z3} \\
&z_{0,3}=1 \overset{g}\longrightarrow z_{3}=\frac{a+b}{c+d}. \notag
\end{align} 
The coefficients $a,b,c,d\in\mathbb{C}$ must also satisfy the condition $ad-bc\neq 0$. These equations make it possible to search for the coefficients of the M\"{o}bius transformation as a function of the fixed points of the orthogonalizing superattractive map we are interested in finding. From Eqs.~(\ref{z1z2z3}) one finds that in order to maintain the orthogonality of the fixed points the relations
\begin{align}
b=&z_{1}d, \notag \\
c=&-z_{1}^{\ast}a, \label{bc}
\end{align}  
must hold between the coefficients of the M\"{o}bius transformation. Thus, by conjugating $f_{0}$ with a M\"{o}bius transformation of the form
\begin{equation}
g(z)=\frac{az+z_{1}d}{-az_{1}^{\ast}z+d}, \,\,\, ad\neq 0, \label{g_ortsup}
\end{equation}
one gets an element of the class of orthogonalizing superattractive nonlinear maps.   

\section{Quantum state matching}
\label{qstateid}

Due to the fact that in the case of the basic transformation $f_{0}$ the Julia set -- the closure of the set of all repelling and neutral fixed cycles -- is a connected set (the unit circle on the complex plane with its center at the origin) and the fact that M\"{o}bius transformations map the fixed points of $f_{0}$ into the fixed points of $f=g\circ f_{0}\circ g^{-1}$ (c.f. Appendix~\ref{appA}), we can determine the Julia set of the new superattractive orthogonalizing nonlinear map $f$ by determining how the unit circle is transformed under the M\"{o}bius transformation $g$.
 
It is a well-known property of M\"{o}bius transformations that they map generalized circles to generalized circles (a generalized circle is either a circle or a line, the latter being considered as a circle through the point at infinity). Therefore, due to the M\"{o}bius transformation $g$ the Julia set of the map $f$ is either a circle or a line on the complex plane. 
The question arises: Is there a way to determine $f$ by defining its Julia set and then finding the M\"{o}bius transformation which maps this circle back into the Julia set of $f_{0}$? This problem would then be equivalent to defining a circle-shaped neighborhood around a complex number $z$ (or around a pure quantum state parameterized by $z$) and determining what nonlinear map $f$ has such convergence property. Then using $f$ iteratively on qubits from an ensemble each being in the same pure quantum state $\ket{\psi}$ one could decide in a few steps whether $\ket{\psi}$ was in the predefined neighborhood or not, since after a few steps of the protocol $\ket{\psi}$ would either be transformed into the desired state (represented by one of the fixed points of $f$) or into the state which is orthogonal to the desired state (represented by the other superattractive fixed point). 

\begin{figure}[tbh]
\includegraphics[width=0.49\columnwidth]{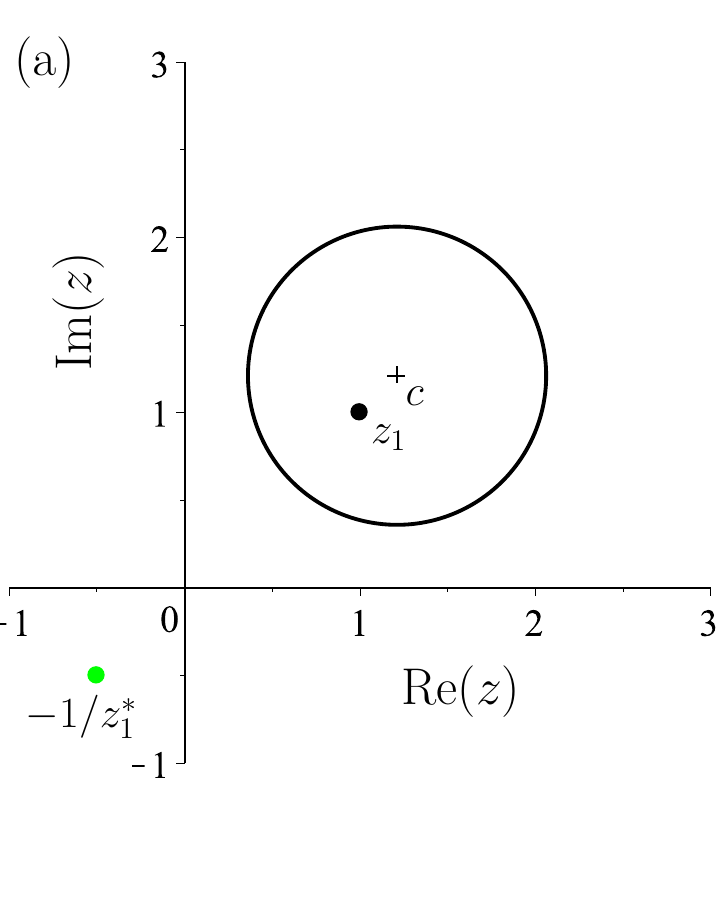}
\includegraphics[width=0.49\columnwidth]{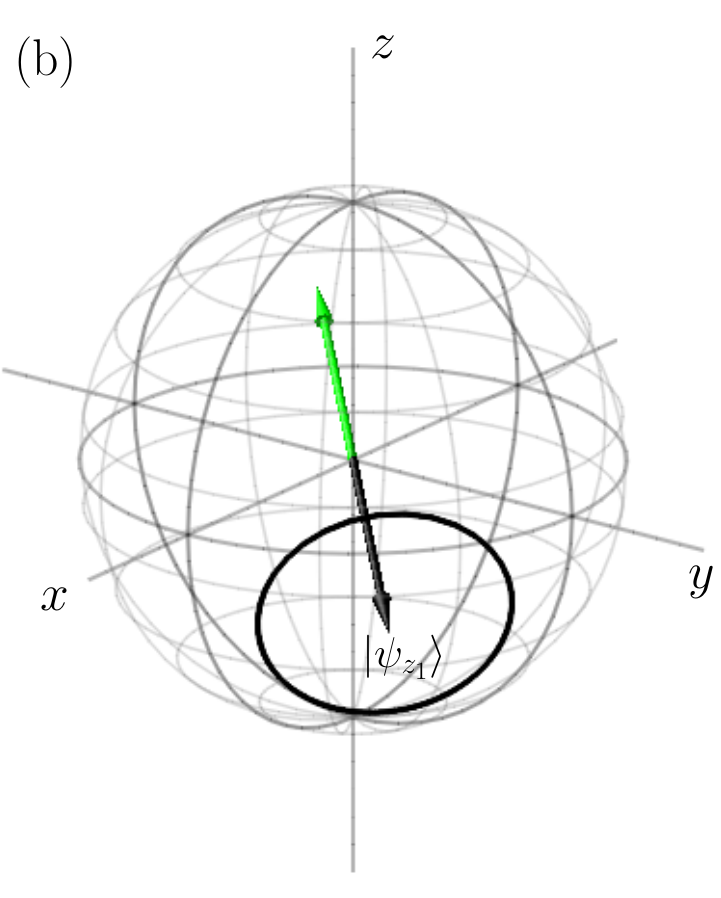}
\caption{(color online) (a) The black circle represents an overlap of $\left|s\right|=0.97$ with the pure state $\ket{\psi_{z_{1}}}$ where $z_{1}=1+i$ (black dot). The point $-1/z_{1}^{\ast}$ corresponding to the state orthogonal to $\ket{\psi_{z_{1}}}$ is shown by the green dot. (b)  Bloch-sphere representation of the scenario of (a). The black circle corresponds to an  overlap of $\left|s\right|=0.97$ with the state $\ket{\psi_{z_{1}}}$. Black and green arrows correspond to states $\ket{\psi_{z_{1}}}$, and $\ket{\psi_{-1/z_{1}^{\ast}}}$, respectively.   
}
\label{Fig1}
\end{figure}

Let us think of the desired pure quantum state as the one which corresponds to the fixed point $z_{1}$ of the yet unknown nonlinear map $f$, and let us define the Julia set of $f$ as a circle of radius $r$ on the complex plane with its center in $c$, i.e., ${\cal{J}}_{f}=\left\lbrace c+re^{i\varphi}, \quad c\in\mathbb{C}, \quad \varphi\in \left[0,2\pi\right)\right\rbrace$. We are looking for the inverse of the M\"{o}bius transformation (\ref{g_ortsup}) which maps this circle back into the unit circle at the origin (while also mapping $z_{1}$ and its orthogonal pair $-1/z_{1}^{\ast}$ into $0$ and $\infty$, respectively). The inverse M\"{o}bius transformation $g^{-1}$ in this case can be written as
\begin{equation}
g^{-1}(z)=\frac{d\left(z-z_{1}\right)}{a\left(1+z_{1}^{\ast}z\right)}.
\end{equation}
Applying the requirement that $g^{-1}$ transforms ${\cal J}_{f}$ into the unit circle ${\cal J}_{f_{0}}$, which can be written as  
\begin{equation}
\left|g^{-1}(z)\right|^{2}=1, \quad z\in{\cal{J}}_{f}
\end{equation}
one finds that 
\begin{equation}
\frac{\left|d\right|^{2}}{\left|a\right|^{2}}=
\frac{\left|1+z_{1}^{\ast}z\right|^{2}}{\left|z-z_{1}\right|^{2}}=
\frac{\left(1+\left|z_{1}\right|^{2}\right)\left(1+\left|z\right|^{2}\right)}{\left|z-z_{1}\right|^{2}}-1
\label{dpa1}
\end{equation}
or equivalently
\begin{equation}
\frac{\left|z-z_{1}\right|^{2}}{\left(1+\left|z_{1}\right|^{2}\right)\left(1+\left|z\right|^{2}\right)}=
\frac{1}{1+\frac{\left|d\right|^{2}}{\left|a\right|^{2}}}=\mathrm{const.} \quad  (z\in{\cal{J}}_{f}).
\end{equation}
Looking at the left-hand side of the above equation it can be easily seen that 
\begin{equation}
\frac{\left|z-z_{1}\right|^{2}}{\left(1+\left|z_{1}\right|^{2}\right)\left(1+\left|z\right|^{2}\right)}=
1-\left|s\right|^{2}, \label{1ms2}
\end{equation}
where $s$ is the scalar product
\begin{equation}
s=\braket{\psi_{z_{1}}}{\psi_{z}}=\frac{1+z_{1}^{\ast}z}
{\sqrt{\left(1+\left|z_{1}\right|^{2}\right)\left(1+\left|z\right|^{2}\right)}}.
\label{scalprod}
\end{equation}
This indicates that for the pure quantum states which correspond to the complex numbers $z\in{\cal{J}}_{f}$ the scalar product with the pure state corresponding to the complex number $z_{1}$ (the fixed point of $f$) is constant. Or conversely, the complex numbers which describe quantum states that have the same overlap with the quantum state $\ket{\psi_{z_{1}}}$ lie on a circle in the complex plane (see Fig.~\ref{Fig1}a). We note that this latter is a general statement, irrespective of $z_{1}$ being a fixed point or not: complex numbers which correspond to a fixed absolute value of the scalar product with a given state described by $z_{1}$ lie on a circle in the complex plane. 

The picture is very intuitive if one represents the pure quantum states $\ket{\psi_{z}}$ and $\ket{\psi_{z_{1}}}$ on the Bloch sphere, where the Bloch vector of pure states with a given overlap (absolute value of the scalar product) with $\ket{\psi_{z_{1}}}$ draw a circle on the surface of the Bloch sphere, and the Bloch vector of $\ket{\psi_{z_{1}}}$ points into the center of this circle, see Fig.~\ref{Fig1}b. On the complex plane, however, this picture is slightly more complicated. Even though a circle corresponds to a fixed absolute value of the scalar product with $\ket{\psi_{z_{1}}}$, the center of the circle is not in $z_{1}$ (see Fig.~\ref{Fig1}a). It is interesting to note that $z_{1}$ is not necessarily inside the circle corresponding to a fixed $\left|s\right|$, as we show in Appendix~\ref{appB}, where we also determine the radius and center of the circle which corresponds to a given $\left|s\right|$ on the complex plane and analyze this phenomenon in more detail.

%%%%%%%%%%%%%%%%%%%%%%%%%%%%%%%%%%%%%%%%%%%%%%%%%%%%%%%%%%%%%%%%%%%%%%%%%%%%%%%% 
%%%%%%%%%%%%%%%%%%%%%%%%%%%%%%%%%%%%%%%%%%%%%%%%%%%%%%%%%%%%%%%%%%%%%%%%%%%%%%%%

In the light of Eq.~(\ref{1ms2}) it can be seen that one may think of defining the Julia set ${\cal{J}}_{f}$ of $f$ as the circle which corresponds to a given minimum absolute value of the scalar product $\left|s_{\varepsilon}\right|=\left|\braket{\psi_{z_{1}}}{\psi_{z}}\right|_{\mathrm{min}}$ with the desired pure state corresponding to the superattractive fixed point $z_{1}$ of $f$. Then, every complex number $z$ which corresponds to an overlap of $\left|s\right|>\left|s_{\varepsilon}\right|$ will converge to $z_{1}$ (or equivalently every $\ket{\psi_{z}}$ will converge to $\ket{\psi_{z_{1}}}$) after a few iterations of $f$ (obviously, states, for which $\left|s\right|<\left|s_{\varepsilon}\right|$ will converge to the state $\ket{\psi_{-1/z_{1}^{\ast}}}$ orthogonal to $\ket{\psi_{z_{1}}}$). Therefore, the protocol which implements $f$ could be used for matching a pure quantum state which is within the neighborhood represented by the circle corresponding to $\left|s_{\varepsilon}\right|$ around $\ket{\psi_{z_{1}}}$ with $\ket{\psi_{z_{1}}}$ by iterating the scheme for many times. Then a projective measurement can decide between the two orthogonal states. Thus, the proposed procedure is an unambiguous scheme in the limit of many iterations. In practice, as we will show in the next section (see Fig.~\ref{Fig2}), the nonlinearity of the process quickly enhances the overlap of the matched state with the reference state. After such a matching protocol has yielded the result that the state of the ensemble is within the prescribed neighborhood of $\ket{\psi_{z_{1}}}$, remaining qubits of the ensemble could be used for quantum state error correction after implementing the same number of iterative steps on them. Thus, in a quantum computational scenario, after correcting their states, these qubits can be further processed in a subsequent computation. In what follows we show how such nonlinear $f$ transformations can be determined for a given desired reference state and a prescribed neighborhood.

If one fixes $\left|s_{\varepsilon}\right|$ then, according to Eqs.~(\ref{dpa1}) and (\ref{1ms2}), the ratio $\left|d\right|\!/\!\left|a\right|$ can be determined as a function of $\left|s_{\varepsilon}\right|$:
\begin{equation}
\frac{\left|d\right|}{\left|a\right|}=\frac{\left|s_{\varepsilon}\right|}{\sqrt{1-\left|s_{\varepsilon}\right|^{2}}}.
\end{equation} 
Therefore, $g^{-1}$ can be written as
\begin{equation}
g^{-1}(z)=\frac{\left|s_{\varepsilon}\right|}{\sqrt{1-\left|s_{\varepsilon}\right|^{2}}}
e^{-i\alpha}\frac{z-z_{1}}{1+z_{1}^{\ast}z},
\end{equation}     
where $\alpha$ is the phase difference of the coefficients $d$ and $a$ (we show later that $\alpha$ may be chosen arbitrarily). 

One may decompose $g^{-1}$ into two simpler M\"{o}bius transformations as
\begin{equation}
g^{-1}=g_{\varepsilon}^{-1}\circ g_{U_{z_{1}}}^{-1},
\end{equation}
where 
\begin{equation}
g_{\varepsilon}^{-1}(z)=\frac{z}{\varepsilon }, \label{gepsm1}
\end{equation}
with 
\begin{equation}
\varepsilon=\left|\varepsilon\right|e^{i\alpha_{\varepsilon}}, \qquad \left|\varepsilon\right|=\frac{\sqrt{1-\left|s_{\varepsilon}\right|^{2}}}{\left|s_{\varepsilon}\right|},
\label{epsilon}
\end{equation}   
and 
\begin{equation}
g_{U_{z_{1}}}^{-1}(z)=e^{-i\alpha_{u}}\frac{z-z_{1}}{1+z_{1}^{\ast}z}, \label{gUm1}
\end{equation}
where $\alpha_{u}+\alpha_{\varepsilon}=\alpha$. Then the M\"{o}bius transformation $g$ can be written as
\begin{equation}
g=\left(g_{\varepsilon}^{-1}\circ g_{U_{z_{1}}}^{-1}\right)^{-1}=g_{U_{z_{1}}}\circ g_{\varepsilon}, \label{gdecomp}
\end{equation}
where $g_{\varepsilon}$ and $g_{U_{z_{1}}}$ are the inverses of (\ref{gepsm1}) and (\ref{gUm1}), respectively
\begin{align}
g_{\varepsilon}(z)&=\varepsilon z \label{g_eps} \\
g_{U_{z_{1}}}(z)&=\frac{e^{i\alpha_{u}}z+z_{1}e^{-i\alpha_{u}}}{-z_{1}^{\ast}e^{i\alpha_{u}}z+e^{-i\alpha_{u}}}. \label{g_Uz1}
\end{align} 

The transformation $g_{\varepsilon}(z)=\varepsilon z$ is an elementary M\"{o}bius transformation which when acting on the Julia set ${\cal{J}}_{f_{0}}$ it changes its radius to $\left|\varepsilon\right|$, but it does not change the superattractive fixed points ($0$ and $\infty$) of $f_{0}$. The phase $\alpha_{\varepsilon}$ does not affect the Julia set as it only causes a phase shift by $\alpha_{\varepsilon}$ on the points belonging to ${\cal{J}}_{f_{0}}$. Thus, $\alpha_{\varepsilon}$ may be chosen arbitrarily.

The other M\"{o}bius transformation $g_{U_{z_{1}}}$ belongs to a special subclass of M\"{o}bius transformations which preserve the "orthogonal" property of the fixed points, while leaving the absolute value of the scalar product (the overlap of the states corresponding to the points of the Julia set with $\ket{\psi_{z_{1}}}$) unchanged. We call these here unitary M\"{o}bius transformations as it can be shown that such M\"{o}bius transformations correspond to unitary single-qubit operations. A general unitary M\"{o}bius transformation can be written in the form
\begin{equation}
g_{U}=\frac{p z+q}{-q^{\ast}z+p^{\ast}}, \quad \text{with} \left|p\right|^{2}+\left|q\right|^{2}=1.
\end{equation}
Comparing $g_{U_{z_{1}}}$ with $g_{U}$ it can be seen that $g_{U_{z_{1}}}$ is indeed a unitary M\"{o}bius transformation with coefficients
\begin{align}
p&=\frac{e^{i\alpha_{u}}}{\sqrt{1+\left|z_{1}\right|^{2}}}, \notag \\
q&=\frac{z_{1}e^{-i\alpha_{u}}}{\sqrt{1+\left|z_{1}\right|^{2}}}.
\end{align}
According to Eq.~(\ref{gdecomp}) $g$ can be decomposed into the subsequent actions of $g_{\varepsilon}$ and $g_{U_{z_{1}}}$. It is $g_{\varepsilon}$ which acts first, and we have seen that it does not change the superattractive fixed points $0$ and $\infty$, it only changes the radius of the Julia set and this change can be identified with an increase or decrease of the overlap of the states corresponding to the Julia set with the state corresponding to the superattractive fixed point $0$. Then we act with $g_{U_{z_{1}}}$, which does not change the overlap, since $g_{U_{z_{1}}}$ is a unitary M\"{o}bius transformation, but changes the superattractive fixed point $0$ to $z_{1}$, and $\infty$ to $-1/z_{1}^{\ast}$, irrespective of the phase $\alpha_{u}$. Since $\alpha_{u}$ does not change the Julia set either (its effect is only a $2\alpha_{u}$ phase shift on the points) it can be chosen arbitrarily. Therefore $\alpha$ is also arbitrary. In what follows we choose $\alpha_{u}=\alpha_{\varepsilon}=\alpha=0$.

\section{Realization of maps suitable for quantum state matching}
\label{realization}

In the previous section we have shown how one can construct a M\"{o}bius transformation $g$ with which, when conjugating the simplest orthogonalizing superattractive nonlinear map $f_{0}$ will produce another orthogonalizing superattractive map $f$. The fixed points of $f$ represent orthogonal pure states and its Julia set corresponds to a minimal overlap with some desired reference state that is represented by one of its fixed points. In order to determine $f$, the following steps need to be followed: (i) set the desired reference state which determines the fixed point $z_{1}$ of $f$ and consequently, $g_{U_{z_{1}}}$ through Eq.~(\ref{g_Uz1}) (ii) set the neighborhood, i.e., the minimal acceptable overlap $\left|s_{\varepsilon}\right|$ with the desired reference state $\ket{\psi_{z_{1}}}$, this determines the Julia set of $f$ and consequently, $g_{\varepsilon}$ through Eq.~(\ref{g_eps}). Then, using $g_{\varepsilon}$ and $g_{U_{z_{1}}}$ to construct $g$ (see Eq.~(\ref{gdecomp})), the nonlinear map can be given as
\begin{equation}
f(z)=g\circ f_{0}\circ g^{-1}=g_{U_{z_{1}}}\circ g_{\varepsilon}\circ f_{0} \circ g_{\varepsilon}^{-1}\circ g_{U_{z_{1}}}^{-1}. \label{fconstr}
\end{equation}
This orthogonalizing superattractive nonlinear map can be used to match pure quantum states with $\ket{\psi_{z_{1}}}$ if they have an initial overlap with $\ket{\psi_{z_{1}}}$ larger than $\left|s_{\varepsilon}\right|$. Now the question is how to construct an appropriate two-qubit gate and a measurement protocol which is able to implement $f$? 

\subsection{Direct approach to implement $f$}
\label{directreal}

Substituting the general forms of $g_{\varepsilon}$ and $g_{U_{z_{1}}}$ given by Eqs.~(\ref{g_eps}) and (\ref{g_Uz1}) with $\alpha_{u}=\alpha_{\varepsilon}=0$ into Eq.~(\ref{fconstr}), $f$ can be written as
\begin{equation}
f(z)\!=\!\frac{\left(\varepsilon z_{1}^{\ast}\left|z_{1}\right|^{2}+1\right)\!z^{2}
+2z_{1}\left(\varepsilon z_{1}^{\ast}-1\right)z
+z_{1}\left(\varepsilon+z_{1}\right)}
{z_{1}^{\ast}\left(\varepsilon z_{1}^{\ast}-1\right)z^{2}
+2z_{1}^{\ast}\left(\varepsilon+z_{1}\right)z
+\varepsilon-z_{1}\left|z_{1}\right|^{2}}.
\label{fgeneral}
\end{equation}   
We note that the form of the two-qubit unitary which can implement $f$ depends on the choice of the  post-selection method. Here we restrict ourselves to the following measurement protocol: after the two-qubit gate acting on qubits $A$ and $B$, we keep qubit $A$ only if the measurement in the $\left\lbrace \ket{0}, \ket{1} \right\rbrace$ basis on qubit $B$ yielded the result $0$. Then, we require that the state of qubit $A$ be transformed into $\ket{0}_{A}+f(z)\ket{1}_{A}$, and look for a two-qubit operation $U$ which corresponds to this scenario. In Appendix~\ref{appC} we show how one can construct an appropriate $U$ for any quadratic rational  map $f$ in the case of the above mentioned measurement protocol (we note that it is then straightforward to construct $U$ for a different protocol).

%%%%%%%%%%%%%%%%%%%%%%%%%%%%%%%%%%%%%%%%%%%%%%%%%%%%%%%%%%%%%%%%%%%%%%%%%%%%%%%%
\begin{figure}[tbh]
\includegraphics[width=0.99\columnwidth]{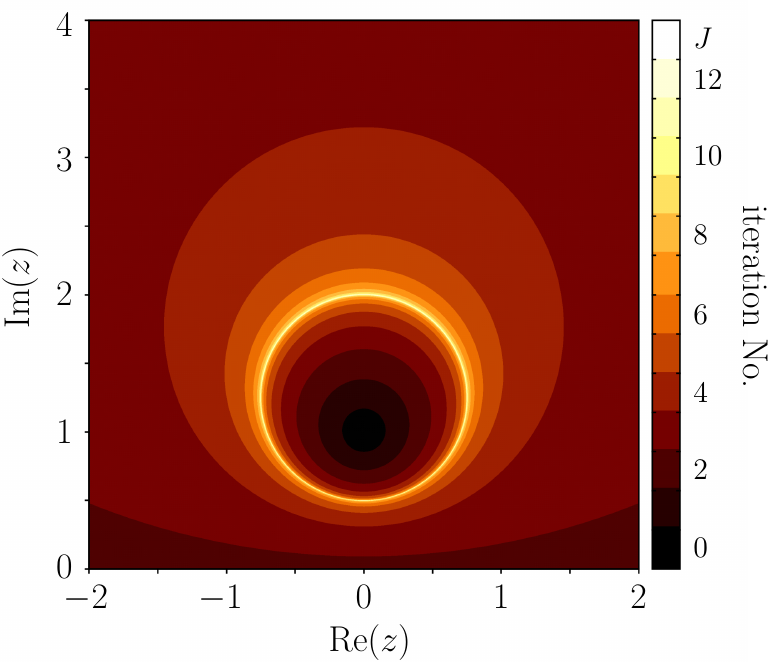}
\caption{(color online) The complex plane after iterations of the nonlinear map which matches qubit states with the state $\ket{\psi_{z_{1}}}=\left(\ket{0}+i\ket{1}\right)/\sqrt{2}$ if they have an initial overlap larger than $\left|s_{\varepsilon}\right|^{2}=0.9$ (which corresponds to $\varepsilon=1/3$). The color code represents the number of iterations needed for a given pure state represented by the respective complex number $z$ to reach an overlap larger than $\left|s\right|^{2}=0.994$ with $\ket{\psi_{z_{1}}}$ or its orthogonal pair $\ket{\psi_{-1/z_{1}^{\ast}}}$. The very thin white circle corresponds to the Julia set, which separates the two convergence regions of the nonlinear map. Quantum states represented by the inner (outer) region of this circle approach $\ket{\psi_{z_{1}}}$ ($\ket{\psi_{-1/z_{1}^{\ast}}}$). The black circle corresponds to states which initially have an overlap larger than $0.994$ with $\ket{\psi_{z_{1}}}$.}
\label{Fig2}
\end{figure}
%%%%%%%%%%%%%%%%%%%%%%%%%%%%%%%%%%%%%%%%%%%%%%%%%%%%%%%%%%%%%%%%%%%%%%%%%%%%%%%%

For the example shown in Fig.~\ref{Fig2}, where the nonlinear map matches qubit states with the state $\ket{\psi_{z_{1}}}=\left(\ket{0}+i\ket{1}\right)/\sqrt{2}$ if they have an initial overlap larger than $\left|s_{\varepsilon}\right|^{2}=0.9$ (which corresponds to $\varepsilon=1/3$), $U$ may be given as  
\begin{equation}
U=\frac{1}{6\sqrt{2}}\left( 
\begin{array}{cccc}
1-3i & 7-i & -1-i & -1+3i \\
6 & 0 & 0 & 6 \\
-3+i & 1+i & 1-7i & 3-i \\
-4 & 2+4i & -2+4i & 4 
\end{array} \right).
\end{equation}
(We note that the second and fourth rows of $U$ can be chosen arbitrarily as long as they form an orthonormal set with the first and third rows, which are determined by the quadratic rational map $f$.) Fig.~\ref{Fig2} shows that in the convergence region which is of interest (inner region of the white circle), quantum states approach $\ket{\psi_{z_{1}}}$ with the given precision after already a few steps of the iteration. Quantum states which have an initial overlap with $\ket{\psi_{z_{1}}}$ that is close to the prescribed value $\left|s_{\varepsilon}\right|$ are harder to be matched with $\ket{\psi_{z_{1}}}$ as it takes more iterations for them to approach the reference state.

\subsection{Implementation of $f$ with a single-qubit gate and a special two-qubit gate}
\label{specialreal}

Eq.~(\ref{fconstr}) suggests that since $g$ can be decomposed into the subsequent action of two simple M\"{o}bius transformations, one of which ($g_{U_{z_{1}}}$) corresponds to a single-qubit unitary operation, it may be possible to implement $f$ as a combination of a single-qubit gate realizing $U_{z_{1}}$ and a two-qubit gate which implements the nonlinear map $f_{\varepsilon}=g_{\varepsilon}\circ f_{N} \circ g_{\varepsilon}^{-1}$. From the point of view of possible applications only the $\left|\varepsilon\right|<1$ case is of interest, i.e., the "contraction" of the Julia set of $f_{0}$ into the required neighborhood. (We note that since $g_{\varepsilon}$ does not correspond to a unitary single-qubit operation, its direct implementation is not possible). 

Using the results of Appendix~\ref{appC}, we can determine a two-qubit unitary matrix which, together with the above mentioned measurement protocol, realizes the nonlinear map 
\begin{equation}
f_{\varepsilon}(z)=g_{\varepsilon}\circ f_{N} \circ g_{\varepsilon}^{-1}=\frac{z^{2}}{\varepsilon}, \qquad \varepsilon=\left|\varepsilon\right|<1,
\label{f_eps}
\end{equation}
where we have chosen $\alpha_{\varepsilon}=0$.
The unitary which implements $f_{\varepsilon}$ may be given as:
\begin{equation}
U_{\varepsilon}=\left( 
\begin{array}{cccc}
\varepsilon & \frac{1}{\sqrt{2}}\sqrt{1-\varepsilon^{2}} & -\frac{1}{\sqrt{2}}\sqrt{1-\varepsilon^{2}} & 0 \\
0 & \frac{1}{\sqrt{2}} & \frac{1}{\sqrt{2}} & 0 \\
0 & 0 & 0 & 1 \\
\sqrt{1-\varepsilon^{2}} & -\frac{1}{\sqrt{2}}\,\varepsilon & \frac{1}{\sqrt{2}}\,\varepsilon & 0
\end{array} \right).
\end{equation} 
(We note again that the second and fourth rows of $U$ can be chosen differently).

\begin{figure}[tbh]
\includegraphics[width=0.49\columnwidth]{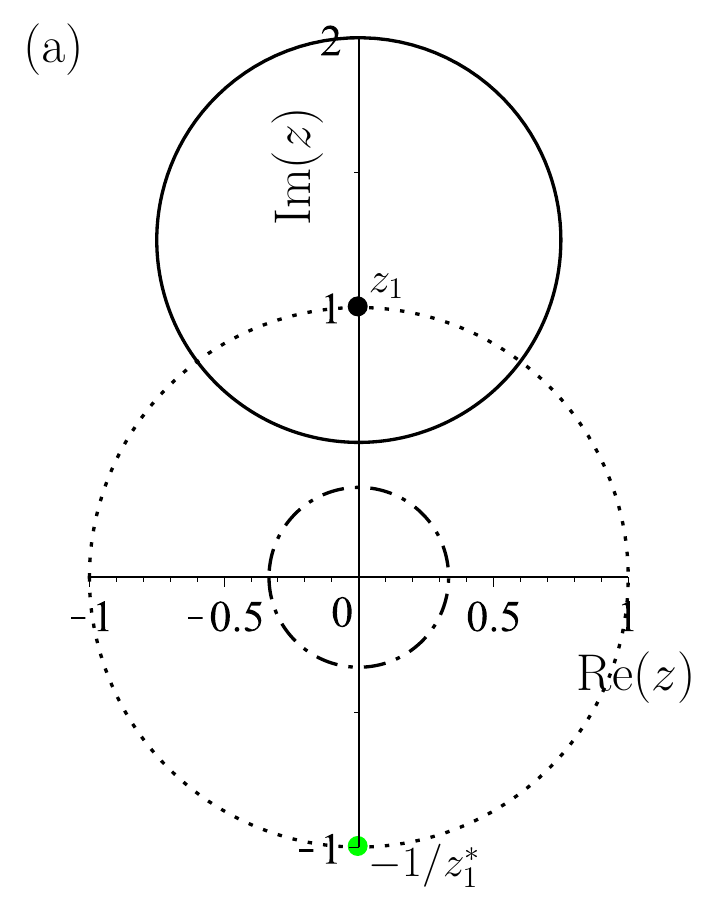}
\includegraphics[width=0.49\columnwidth]{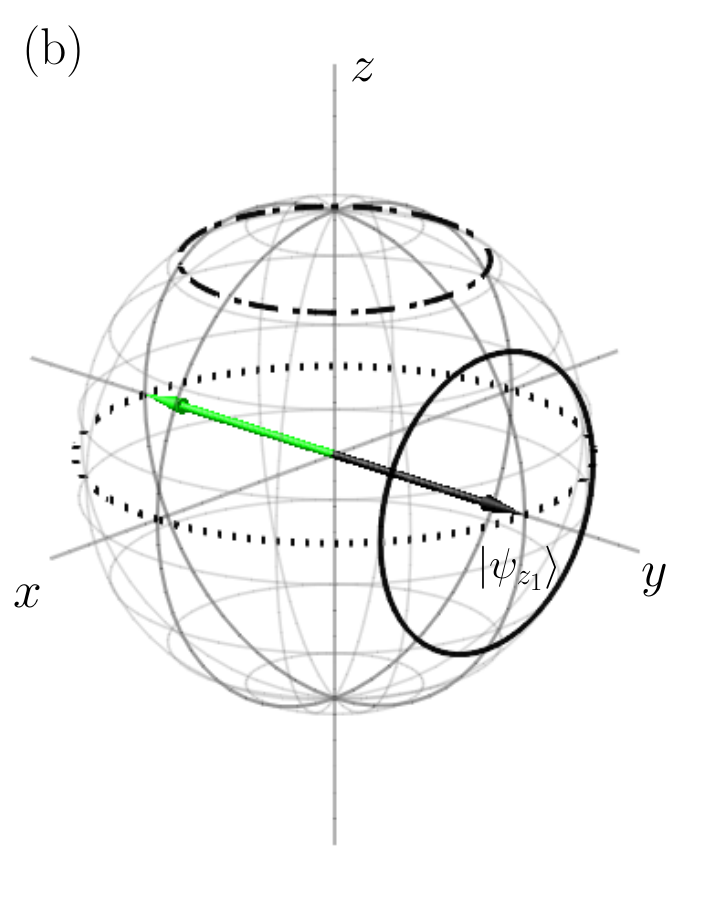}
\caption{(color online) The decomposition of the map $f$ into the subsequent actions of a contracting two-qubit operation and a single-qubit unitary operation. The same parameters have been used as for Fig.~\ref{Fig2}. (a) The effect of the decomposition represented on the complex plane. The original Julia set (that of $f_{0}$) is shown by a dotted circle, which is then contracted by $\varepsilon=1/3$ into the Julia set of $f_{\varepsilon}$ (dash-dotted circle). The final Julia set (although corresponds to the same $\left|s_{\varepsilon}\right|$ as that of $f_{\varepsilon}$) appears to be enlarged (black solid circle). (b) The decomposition represented on the Bloch sphere. The contracted Julia set (dash-dotted circle) is rotated by the unitary single-qubit operation into the black circle, while its radius is unchanged. 
}
\label{Fig3}
\end{figure}

Fig.~\ref{Fig3} shows how the implementation of $f$ (for the same parameters as in the case of Fig.~\ref{Fig2}) can be decomposed into the subsequent actions of a contracting two-qubit operation and a single-qubit unitary operation (Fig.~\ref{Fig3}a shows this decomposition  on the complex plane, while Fig.~\ref{Fig3}b shows this on the Bloch sphere). Note that the solid black circle in Fig.~\ref{Fig3}a (the Julia set of $f$ which has been identified with the neighborhood corresponding to the overlap $\left|s_{\varepsilon}\right|$) is similar to the white circle in Fig.~\ref{Fig2}. The effect of $g_{\varepsilon}$ is clearly a contraction of the Julia set, both on the complex plane and on the Bloch sphere. Interestingly, although the final Julia set corresponds to the same $\left|s_{\varepsilon}\right|$ as in the contracted case, it appears to be enlarged when represented on the complex plane (solid circle in Fig.~\ref{Fig3}a). When represented on the Bloch sphere (Fig.~\ref{Fig3}b), it can be clearly seen that the Julia set of the contracting operation is in fact not deformed after the action of the single-gubit unitary operation which only rotates the contracted Julia set (dash-dotted cirlce) into a circle the center of which is the desired reference state $\ket{\psi_{z_{1}}}$ (solid circle).

%%%%%%%%%%%%%%%%%%%%%%%%%%%%%%%%%%%%%%%%%%%%%%%%%%%%%%%%%%%%%%%%%%%%%%%%%%%%%%%% 
%%%%%%%%%%%%%%%%%%%%%%%%%%%%%%%%%%%%%%%%%%%%%%%%%%%%%%%%%%%%%%%%%%%%%%%%%%%%%%%%

\subsection{Success probability}
\label{succprob}

The success probability of a single step of the protocol is the same for the two types of implementations presented above. Following the decomposition of $f$ into a single-qubit rotation and the subsequent application of the unitary $U_{\varepsilon}$ (case of Sec.~\ref{specialreal}) it is easy to see that the success probability of measuring qubit $B$ in the state $\left|0\right>$ can be given as
\begin{equation}
{\cal P}_{\varepsilon}
=\frac{\varepsilon^{2}}{\left(1+|g_{U_{z_{1}}}^{-1}|^{2}\right)^{2}}
\left(1+\left|f_{\varepsilon}\left(g_{U_{z_{1}}}^{-1}\right)\right|^{2}\right).
\end{equation}
Using Eqs. (\ref{1ms2}), (\ref{epsilon}) and (\ref{gUm1}) this can be written in the simple form
\begin{equation}
{\cal P}_{\varepsilon}=\varepsilon^{2}\left|s\right|^{4}+\left(1-\left|s\right|^{2}\right)^{2},
\end{equation}
where $\left|s\right|^{2}$ is the square of the absolute value of the scalar product of the initial state $\left|\psi_{z}\right>$ with the reference state $\left|\psi_{z_{1}}\right>$, while $\varepsilon$ corresponds to the prescribed neighborhood (the circle which is described by the $\left|s_{\varepsilon}\right|$ scalar product). States with a given overlap with the reference state (which lie on a circle on the Bloch sphere as well as on the complex plane) result in the same success probability. 

%%%%%%%%%%%%%%%%%%%%%%%%%%%%%%%%%%%%%%%%%%%%%%%%%%%%%%%%%%%%%%%%%%%%%%%%%%%%%%%% 
\begin{figure}[tbh]
\includegraphics[width=0.99\columnwidth]{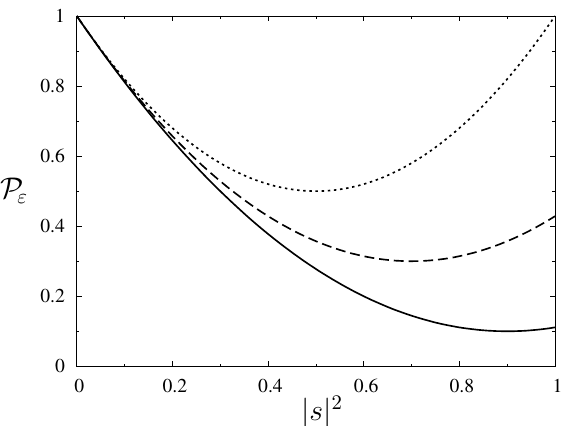}
\caption{ The success probability ${\cal P}_{\varepsilon}$ as a function of the square of the absolute value of the scalar product of the unknown state with the reference state for $\varepsilon=1$ (dotted curve), $\sqrt{3/7}$ (dashed curve), and $1/3$ (solid curve). The minima of the curves are located at the scalar product corresponding to the prescribed neighborhood defined by $\left|s_{\varepsilon}\right|^{2}=0.5$, $0.7$, and $0.9$, respectively. Note that the solid curve corresponds to the case that was considered in Figs.~\ref{Fig2} and \ref{Fig3}.   
}
\label{Fig4}
\end{figure}
%%%%%%%%%%%%%%%%%%%%%%%%%%%%%%%%%%%%%%%%%%%%%%%%%%%%%%%%%%%%%%%%%%%%%%%%%%%%%%%%

Fig.~\ref{Fig4} illustrates ${\cal P}_{\varepsilon}$ as a function of $\left|s\right|^{2}$ for different values of $\varepsilon$: If the prescribed neighborhood is the great circle which is at equal distance from the reference state and its orthogonal pair ($\varepsilon=1$), then the success probability behaves symmetrically inside the two hemispheres of the Bloch sphere. However, if the circle is contracted towards the reference state ($\varepsilon<1$) then ${\cal P}_{\varepsilon}$ becomes asymmetric with respect to the circle: the more one contracts the neighborhood the less the success probability becomes inside it. This means that the stricter the conditions are for a given unknown state to be matched with the reference state, the more resources we need in order to perform the quantum state matching protocol. We note, however, that ${\cal P}_{\varepsilon}$ increases in every subsequential step as the protocol transforms the unknown state closer and closer to the reference state (or its orthogonal pair). 

\subsection{Initial noise} 
\label{initnoise} 

So far we have assumed that there is an ensemble of qubits in the same pure state which we want to match with some reference pure state. What happens if there is some statistical noise in the initial state of the qubits of the ensemble? This scenario can be described by considering mixed initial states for the above defined protocol. For the sake of simplicity, here we confine ourselves to the analysis of the problem in the case of the contracting map $f_{\varepsilon}$ defined by Eq. (\ref{f_eps}), since every quantum state matching nonlinear transformation can be decomposed into the subsequent application of a single-qubit rotation and an $f_{\varepsilon}$ (with a given $\varepsilon$). This means that all $f$ which can be decomposed into the same $f_{\varepsilon}$ have the same convergence properties, but rotated on the Bloch sphere according to the the single-qubit rotation in their decomposition. 

\begin{figure}[tbh]
\includegraphics[height=3.6cm]{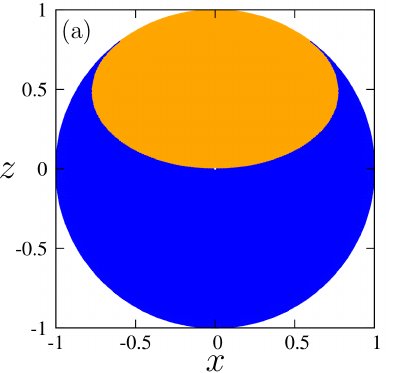}
\includegraphics[height=3.6cm]{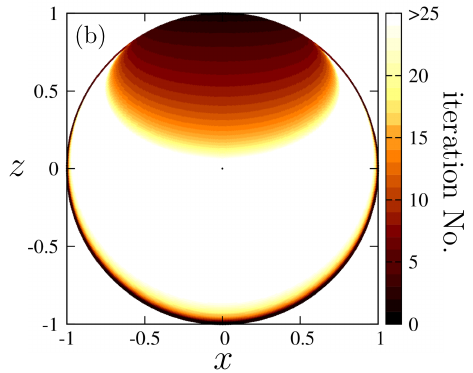}
\caption{(color online) (a) The convergence regions of the transformation $f_{\varepsilon}$ for the same value of $\varepsilon$ as in Fig.~\ref{Fig2}, represented on the $x-z$ plane of the Bloch sphere for initally mixed states. Initial mixed states in the orange (blue) region converge to the $\left|0\right>$ ($\left|1\right>$) pure state. The completely mixed state, represented by the white dot, corresponds to an unstable fixed point of the map $f_{\varepsilon}$. Note that the points where the two convergence regions meet on the surface of the Bloch sphere correspond to the "pure-state neighborhood" which is fixed by $\varepsilon$. (b) The number of iterational steps needed to reach the pure states $\left|0\right>$ or $\left|1\right>$ in the different regions. Here, thecompletely mixed state which is left unchanged by the transformation, is represented by the black dot.}
\label{Fig5}
\end{figure}

By applying $f_{\varepsilon}$ on mixed states it can be shown that the diagonal elements as well as the absolute value of the off-diagonal elements of the resulting density matrix depend only on those of the previous iterational step. The phase of the off-diagonal elements doubles in every iterational step. By the iteration of $f_{\varepsilon}$ either the upper diagonal element increases and the lower one decreases (or vice versa), while the absolute value of the off-diagonal elements gradually decreases, thus, eventually, the mixed state is purified into either of the two fixed states $\left|0\right>$ or $\left|1\right>$ (except for the case when the initial density matrix is the complete mixture, which is an unstable fixed point of the transformation). Due to this property, the convergence regions of initial mixed states on any cut of the Bloch sphere which includes the north and south pole (i.e., the states $\left|0\right>$ and $\left|1\right>$) are the same. In Fig.~\ref{Fig5}a we show the $x-z$ plane of the Bloch sphere, orange (blue) representing mixed initial states which are purified into $\left|0\right>$ ($\left|1\right>$). 
It can be seen that the protocol performs essentially the same way as for pure initial states: mixed states converge to the reference state or to its orthogonal pair, however, the border between the two types of states is not a cone underneath the prescribed neighborhood, but a slightly enlarged lentil shaped structure. Furthermore, the number of iterational steps that is needed to reach the final states on the two sides of this border may differ significantly: While for mixed states lying in the blue region, only small mixedness can be purified with a reasonable number of iterational steps, for mixed states from the orange region the protocol can tolerate more mixedness (see Fig.~\ref{Fig5}b). In the case of no a priori information about the statistical distribution of the noise itself, the protocol may struggle to give a conclusive answer in the close proximity of the border, however, it may  perform as well as in the pure-state case for slightly mixed states. On the other hand, if instead of quantum state matching, one is rather aiming at quantum state error correction for such noisy ensembles, in the possession of some a priori information about the distribution of the noise, one may be able to design a nonlinear transformation with which the initial noise can efficiently be reduced.

%%%%%%%%%%%%%%%%%%%%%%%%%%%%%%%%%%%%%%%%%%%%%%%%%%%%%%%%%%%%%%%%%%%%%%%%%%%%%%%% 
%%%%%%%%%%%%%%%%%%%%%%%%%%%%%%%%%%%%%%%%%%%%%%%%%%%%%%%%%%%%%%%%%%%%%%%%%%%%%%%%
 
\section{Conclusion}
\label{concl}
We considered the problem of quantum state matching, namely to decide whether the pure quantum state of an ensemble of qubits is close to a desired pure reference state. 
We have presented a method based on the iterative application of a postselective scheme. The initial qubits are considered pairwise and one of them is measured after an entangling unitary gate has operated on them, while the remaining qubit's state undergoes an effective nonlinear transformation.
We have determined a class of quantum state transformations of this type which orthogonalize quantum states after a few iterations.
The two convergence regions of such a map are separated by a circle on the Bloch sphere where one of the orthogonal states lies at the center of the inner while the other at the center of the outer region of the circle. The circle corresponds to states of fixed overlap with the two orthogonal states.  We have proven that one can construct a nonlinear transformation for any circle with arbitrary center and radius as a convergence region, thereby we demonstrated that these maps may be utilized to solve the problem of  quantum state matching.  

From a practical point of view, the realization of the scheme requires the repeated application of a single unitary two-qubit quantum gate and a single-qubit projective measurement e.g. in the $\ket{0}$ state of the computational basis. The two-qubit unitary gate is specific to the reference state and the radius of the prescribed neighborhood, but it can be decomposed into a universal, contracting, one parameter two-qubit gate and a generic one-qubit gate. 

Let us note that our procedure provides an alternative to usual tomographic quantum state reconstruction techniques \cite{Schmied}. In our case an iterational step sometimes fails and we have to throw away the participating qubits, but if it succeeds then we can be sure that the resulting state of the qubits remains pure and gets deterministically closer to the reference state (or its pair). In this sense, the scheme is closer in spirit to unambiguous state discrimination procedures while the state that can be reconstructed with usual tomographic methods is in general a mixed state with some uncertainty involved \cite{Blume-Kohout}. Here however, the scheme tolerates some initial noise in the unknown quantum state of the ensemble. Moreover, initially mixed states are purified during the process and converge to the reference state or its orthogonal pair.  

Nonlinear quantum evolution would provide us with quick solutions of hard problems \cite{Childs2016}, whereas in standard quantum mechanics nonlinearity may arise in a probabilistic manner, where  the gains of nonlinear evolution are necessarily accompanied by an increasing number of discarded systems \cite{Gilyen2016}. Nevertheless, the iterative type of dynamics considered here has the advantage of requiring only a single apparatus implementing the necessary quantum gates. The same apparatus can be reused in subsequent steps of the iteration. The practical advantage of such schemes is evident when production, storage and refeeding of qubits is easy while building quantum gates is hard. 

\begin{acknowledgments}
This work was supported by the National Research, Development and Innovation Office (Project Nos. K115624, K124351, PD120975, 2017-1.2.1-NKP-2017-00001). O. K. acknowledges support from the J\' anos Bolyai Research Scholarship of the Hungarian Academy of Sciences. 

\end{acknowledgments}

\appendix
\section{Effect of a M\"{o}bius transformation on the fixed points of a nonlinear map} \label{appA}

The effect of a M\"{o}bius transformation $g(z)$ on the fixed points of a quadratic rational map $f(z)$ is most apparent if we decompose  $g(z)=\frac{az+b}{cz+d}$ into a sequence of four simple M\"{o}bius transformations
\begin{equation}
g(z)=\left(g_{4}\circ g_{3} \circ g_{2} \circ g_{1} \right)(z)=\dfrac{az+b}{cz+d},
\end{equation}
with
\begin{align}
g_{1}(z)&=z+\frac{d}{c}, \notag \\
g_{2}(z)&=\frac{1}{z}, \notag \\
g_{3}(z)&=\frac{bc-ad}{c^2}z \\
g_{4}(z)&=z+\frac{a}{c}. \notag
\end{align}
Taking $f_{0}(z)=z^{2}$ as a simple example of a quadratic rational map, it is easy to check that the effect of $g_{1}(z)$ and $g_{4}(z)$ on the fixed points is a translation by $d/c$ and $a/c$, respectively. $g_{2}(z)$ transforms the fixed points into their reciprocal, while $g_{3}(z)$ corresponds to a multiplication of the fixed points by $(bc-ad)/c^2$. Therefore, in general, a M\"{o}bius transformation $g(z)=(az+b)/(cz+d)$ transforms the fixed points $z_{i}$ of a quadratic rational map $f(z)$ as
\begin{equation}
z_{i}\rq=g(z_{i})=\frac{az_{i}+b}{cz_{i}+d}.
\label{fptransf}
\end{equation}

\section{Circle corresponding to a fixed magnitude of the scalar product on the complex plane} \label{appB}

The set of points that correspond to quantum states which have the same $\left|s\right|$ overlap with a given pure state $\ket{\psi_{z_{1}}}$ can be found on a circle in the complex plane. Here we determine the center and the radius of the circle as a function of $z_{1}$ and $\left|s\right|$.

If one takes the square of the absolute value of Eq.~(\ref{scalprod}) and parameterizes $z$ as $z=c+re^{i\varphi}$, then the resulting equation can be used to determine $c$ and $r$:
\begin{multline}
1+z_{1}c^{\ast}+z_{1}^{\ast}c+r\left(z_{1}e^{-i\varphi}+z_{1}^{\ast}e^{i\varphi}\right)\\
+\left|z_{1}\right|^{2}\left[\left|c\right|^{2}+r^{2}+r\left(ce^{-i\varphi}+c^{\ast}e^{i\varphi}\right)\right]
=\\
\left|s\right|^{2}\left(1+\left|z_{1}\right|^{2}\right)\left[1+\left|c\right|^{2}+r^{2}+
r\left(ce^{-i\varphi}+c^{\ast}e^{i\varphi}\right)\right] \notag
\end{multline} 
By substituting e.g. $\varphi=0$, $\pi/2$, and $\pi$ into the above equation one gets a system of three equations which can be solved for $c$ and $r$, giving:
\begin{align}
c&=\frac{z_{1}}{\left|s\right|^{2}\left(1+\left|z_{1}\right|^{2}\right)-\left|z_{1}\right|^{2}},\\
r&=\frac{\left|s\right|\left(1+\left|z_{1}\right|^{2}\right)\sqrt{1-\left|s\right|^{2}}}
{\left|\left|s\right|^{2}\left(1+\left|z_{1}\right|^{2}\right)-\left|z_{1}\right|^{2}\right|}.
\end{align}

If $z_{1}=c=0$, then the points which correspond to states with a given overlap $\left|s\right|$ with the state $\ket{0}$ are on a circle of radius $r=\sqrt{1-\left|s\right|^{2}}/\left|s\right|$.

By calculating $\left|c-z_{1}\right|^{2}$ it is easy to see that when $z_{1}\neq 0$ then $z_{1}$ is inside the circle if $\left|s\right|^{2}>\left|s_{\mathrm{0}}\right|^{2}=\left|z_{1}\right|^{2}/\left(1+\left|z_{1}\right|^{2}\right)$. If $\left|s\right|^{2}=\left|s_{\mathrm{0}}\right|^{2}$ then one gets a line which crosses the origo. Otherwise it is not $z_{1}$ but its orthogonal pair $-1/z_{1}^{\ast}$ which is inside the circle corresponding to a fixed overlap.

Interestingly, $z_{1}$ may be inside the circle, even though the overlap with $z_{1}$ is smaller than the overlap with $-1/z_{1}^{\ast}$, this is the case for $\left|z_{1}\right|^{2}<1$, or $z_{1}$ may be outside the circle, even though the overlap with $z_{1}$ is larger than the overlap with $-1/z_{1}^{\ast}$, which is the case for $\left|z_{1}\right|^{2}>1$. We note that this is a phenomenon specific to representing quantum states on the complex plane. When representing states on the Bloch sphere, a fixed overlap again corresponds to a circle on the surface, but the center of the circle is $z_{1}$ as long as $\left|s\right|^{2}>1/2$ (the $\left|s\right|^{2}=1/2$ case corresponds to a great circle). In the $\left|z_{1}\right|^{2}=1$ case $\left|s_{\mathrm{0}}\right|^{2}=1/2$, therefore an equal overlap with $z_{1}$ and with $-1/z_{1}^{\ast}$ is represented by a straight line in the complex plane. This was the case that was observed in \cite{Torres2017}.

\section{Two-qubit unitary transformation to implement a quadratic rational map}
\label{appC}

Here we give a method to determine a two-qubit unitary transformation $U$ which, together with a properly defined post-selection protocol, implements a quadratic rational map
\begin{equation}
f(z)=\frac{a_{0}z^{2}+a_{1}z+a_{2}}{b_{0}z^{2}+b_{1}z+b_{2}}.
\end{equation}
We note that the method is the two-qubit version of the method presented for an $n$-qubit protocol in Ref.\cite{Gilyen2016}. 

Let us assume that we have two qubits in the product state (\ref{Psi12}) and a two-qubit unitary transformation $U$ which can be represented in the $\left\lbrace\ket{00},\ket{01},\ket{10},\ket{11}\right\rbrace$ basis by a matrix $\left[ u_{j,k}\right]_{j,k=1,..,4}$. We choose the following post-selection protocol: after the action of the two-qubit gate $U$, we measure whether the state of qubit $B$ is $\ket{0}_{B}$. Conditioned on this measurement result, the state of qubit $A$ becomes 
\begin{equation}
\ket{\psi_{1}}_{A}=\frac{1}{\cal N}\left[\ket{0}_{A}+\frac{u_{31}+\left(u_{32}+u_{33}\right)z+u_{34}z^{2}}
{u_{11}+\left(u_{12}+u_{13}\right)z+u_{14}z^{2}}\ket{1}_{A}\right],
\end{equation}   
where ${\cal N}=\sqrt{\braket{\psi_{1}}{\psi_{1}}_{A}}$. It can be seen that it is the first and the third row of the matrix of $U$ which determine the map $f(z)$. Since the rows (and columns) of a unitary matrix form an orthonormal set, we need to find two vectors (say $\ket{\tilde{u}^{(1)}}$ and $\ket{\tilde{u}^{(3)}}$) with the following properties:
\begin{align}
\braket{\tilde{u}^{(3)}}{\tilde{u}^{(1)}}&=0, \label{ort} \\
\braket{\tilde{u}^{(3)}}{\tilde{u}^{(3)}}&=\braket{\tilde{u}^{(1)}}{\tilde{u}^{(1)}}, \label{norm}\\
\tilde{u}_{31}&=a_{2}, \notag \\
\tilde{u}_{34}&=a_{0}, \notag \\
\tilde{u}_{11}&=b_{2}, \notag \\
\tilde{u}_{14}&=b_{0}, \notag \\
\tilde{u}_{32}+\tilde{u}_{33}&=a_{1}, \notag \\
\tilde{u}_{12}+\tilde{u}_{13}&=b_{1}, \notag
\end{align}
and then normalize them. In order to determine $\tilde{u}_{12}$, $\tilde{u}_{13}$, $\tilde{u}_{32}$, and $\tilde{u}_{33}$, let us introduce the variables 
\begin{align}
\tilde{a}&=\tilde{u}_{32}-\tilde{u}_{33}, \notag \\
\tilde{b}&=\tilde{u}_{12}-\tilde{u}_{13}. \notag
\end{align}
Then Eqs.~(\ref{ort}) and (\ref{norm}) become
\begin{align}
\left|\tilde{a}\right|^{2}\!-\!|\tilde{b}|^{2}\!&=\!2\left[\left|b_{2}\right|^{2}
\!+\!\frac{\left|b_{1}\right|^{2}}{2}\!+\!\left|b_{0}\right|^{2}
\!-\!\left|a_{2}\right|^{2}\!-\!\frac{\left|a_{1}\right|^{2}}{2}
\!-\!\left|a_{0}\right|^{2}\right], \notag \\
\tilde{a}^{\ast}\tilde{b}&=-2\left(a_{2}^{\ast}b_{2}+\frac{a_{1}^{\ast}b_{1}}{2}+a_{0}^{\ast}b_{0}\right).
\end{align}
These formulas can be used to determine the components of $\ket{\tilde{u}^{(1)}}$ and $\ket{\tilde{u}^{(3)}}$ for a given $f$. 
Then determining the normalized vectors $\ket{u^{(1)}}=\ket{\tilde{u}^{(1)}}/\left|\left|\tilde{u}^{(1)}\right|\right|$ and $\ket{u^{(3)}}=\ket{\tilde{u}^{(3)}}/\left|\left|\tilde{u}^{(3)}\right|\right|$, one can look for two more normalized vectors $\ket{u^{(2)}}$ and $\ket{u^{(4)}}$ orthogonal to $\ket{u^{(1)}}$ and $\ket{u^{(3)}}$ to constitute the remaining two rows of $U$.

\end{document}